\documentclass[conference]{IEEEtran}
\IEEEoverridecommandlockouts
\usepackage{cite}
\usepackage{amsmath,amssymb,amsfonts}
\usepackage{algorithmic}
\usepackage{graphicx}
\usepackage{textcomp}
\usepackage{xcolor}
\def\BibTeX{{\rm B\kern-.05em{\sc i\kern-.025em b}\kern-.08em
    T\kern-.1667em\lower.7ex\hbox{E}\kern-.125emX}}

\usepackage{pgfplots}
\pgfplotsset{compat=newest}
\pgfplotsset{plot coordinates/math parser=false} 
\usepgfplotslibrary{groupplots}
\usepackage{tikz}
\usetikzlibrary{matrix}
\usetikzlibrary{positioning}
\tikzset{
	position/.style args={#1:#2 from #3}{
		at=(#3.#1), anchor=#1+180, shift=(#1:#2)
	}
}
\newlength\figureheight
\newlength\figurewidth 

\usepackage{hhline}
\usepackage{subfig}

\usepackage{booktabs}

\newcommand\numberthis{\addtocounter{equation}{1}\tag{\theequation}}

\begin{document}

\title{Optimizing Rate-Distortion Performance of Motion Compensated Wavelet Lifting with Denoised Prediction and Update}

\author{\IEEEauthorblockN{Daniela Lanz, Andr\'e Kaup}
\IEEEauthorblockA{\textit{Multimedia Communications and Signal Processing} \\
\textit{Friedrich-Alexander University Erlangen-N\"{u}rnberg (FAU)}\\
Cauerstr. 7, 91058 Erlangen, Germany \\
\{Daniela.Lanz,Andre.Kaup\}@FAU.de}
}

\maketitle

\IEEEpubid{\begin{minipage}[t]{\textwidth}\ \\[10pt] 978-1-7281-9320-5/20/\$31.00~\copyright~2020 IEEE\end{minipage}}
%\IEEEpubid{\makebox[\columnwidth]{978-1-7281-9320-5/20/\$31.00 \copyright 2020 IEEE \hfill} \hspace{\columnsep}\makebox[\columnwidth]{ }}

\begin{abstract}
Efficient lossless coding of medical volume data with temporal axis can be achieved by motion compensated wavelet lifting. As side benefit, a scalable bit stream is generated, which allows for displaying the data at different resolution layers, highly demanded for telemedicine applications. Additionally, the similarity of the temporal base layer to the input sequence is preserved by the use of motion compensated temporal filtering. However, for medical sequences the overall rate is increased due to the specific noise characteristics of the data.
%However, if a temporal layer shall be used for diagnostic purposes, a high similarity to the input sequence is indispensable. Therefore, motion compensated temporal filtering is employed, leading to a high-quality temporal respresenation, but also to an increasing rate when medical data is considered.
The use of denoising filters inside the lifting structure can improve the compression efficiency significantly without endangering the property of perfect reconstruction. However, the design of an optimum filter is a crucial task. In this paper, we present a new method for selecting the optimal filter strength for a certain denoising filter in a rate-distortion sense. This allows to minimize the required rate based on a single input parameter for the encoder to control the requested distortion of the temporal base layer.
\end{abstract}


\section{Introduction}
\begin{figure}
	\centering
	\includegraphics[width=0.98\columnwidth]{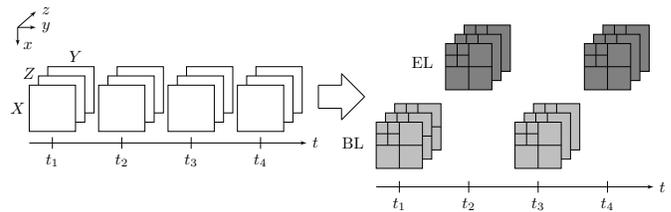}
	\caption{Considered scenario to achieve a fully scalable representation of a 3-D+$t$ volume using wavelet transforms in temporal as well as spatial directions.}
	\label{fig:data}
%	\vspace{-3ex}
\end{figure}

Lossless compression is an important aspect for medical imaging and a challenging requirement for fast transmission at the same time. 
%Considering tasks like remote surgery systems or exchange of medical content between experts, an efficient coding scheme is required.
This becomes even more difficult when high amounts of data are considered, especially dynamic volume data from Computed Tomography (CT). Such 3-D+$t$ data sets consist of $T$ temporally equidistant 3-D volumes of size $X{\times}Y{\times}Z$. On the left side of Fig.~\ref{fig:data}, the basic structure is exemplary shown for $T{=}4$. By recording, e.g., a human thorax over multiple cardiac cycles data rates up to 200\,Mbit/s can be generated~\cite{Ohnesorge2007}. 
So far, medical image compression is mainly performed by the still image coders JPEG~\cite{ITU1992}, JPEG-LS~\cite{ITU2002}, and JPEG2000~\cite{ITU-T-800}, which represent accepted formats of the Digital Imaging and Communications in Medicine~(DICOM) standard~\cite{dicom2012}. 
To achieve further compression gains, temporal redundancies should be exploited.
By splitting the 3-D+$t$ data into $Z$ temporal sequences and handle each sequence separately, the state-of-the-art video codec HEVC/H.265~\cite{2016} can be applied. As a matter of course, the typical bit depth of medical data, comprising 12 bits per sample, as well as the requirement of lossless coding have to be taken into account during coding. Since recently, DICOM provides a mechanism for supporting the use of \mbox{MPEG-2~\cite{ITU1995}}, \mbox{MPEG-4 AVC/H.264~\cite{ITU-T-H-264}}, and \mbox{HEVC/H.265}. However, some health care devices and applications are still lacking in supporting these functionalities~\cite{Shoaib2020}.

An alternative coding scheme is given by \mbox{3D subband coding~(3D-SBC)}~\cite{Karlsson1988}, which is based on discrete wavelet transforms. In \cite{Lanz2018} it has been shown that 3D-SBC outperforms HEVC for lossless compression of noisy CT data. Additionally, 3D-SBC inherently provides different types of scalability, as can be seen in Fig.~\ref{fig:data}.
Besides a lossy base layer with different temporal and spatial resolution, a lossless enhancement layer can be transmitted on demand~\cite{1369697}. Temporal scalability is achieved by performing a 1D wavelet transform in temporal direction, resulting in two temporal subbands. A further decomposition of both subbands by a 2D wavelet transform yields spatial scalability. This can be realized by the JPEG2000 codec, combining spatial and quality scalability as well as the generation of a DICOM conform bit stream in one framework. 
%This is very advantageous for telemedicine applications, where a lossy BL is sufficient for tasks like browsing and fast previewing.
%However, if a diagnosis is exclusively based on remote medical image assessment, the lossless compressed EL can be transmitted if required.
By incorporating motion compensated temporal filtering~(MCTF)~\cite{334985} into the temporal wavelet transform, the fidelity of the base layer can be enhanced. Ghosting artifacts due to temporal displacement in the sequence are reduced by adapting the transform directly to the signal. This is advantageous for telemedicine applications, where the temporal base layer shall be used for diagnostic purposes~\cite{Doukas:2008:ATM:1556147.1362837}.
However, the coding efficiency of noisy CT data suffers from MCTF, which has also been shown in~\cite{Lanz2018}. 
%In a first step a compensated 1-D WT in temporal direction is performed, resulting in two temporal subbands, as can be seen in Fig.~\ref{fig:data}. By further decomposing both subbands by a 2-D WT in spatial direction a fully scalable representation of the original sequence is achieved. 
%This allows for transmitting in a first instance a base layer~(BL) with coarser quality for telemedicine applications like browsing and fast previewing. If required, an enhancement layer~(EL) can be transmitted additionally, which allows for reconstructing the original sequence without any loss.

To overcome this drawback, compensated wavelet lifting with denoised prediction and update~(WLDPU) has recently been introduced~\cite{lnt2019-37}. However, the choice of the optimum filter strength for a specific denoising filter is a crucial task. Therefore, in this work, a rate-distortion optimization\,(RDO) is integrated into the existing framework of WLDPU. 
%This allows for selecting the best matching filter strength for every denoising process.
Section~\ref{sec:WL} will give a short overview of compensated wavelet lifting and WLDPU, followed by a detailed description of the proposed method. Section~4 will give some simulation results, while the last section summarizes this work.

\section{Compensated Wavelet Lifting}
\label{sec:WL}

\begin{figure}
	\includegraphics[width=0.98\columnwidth]{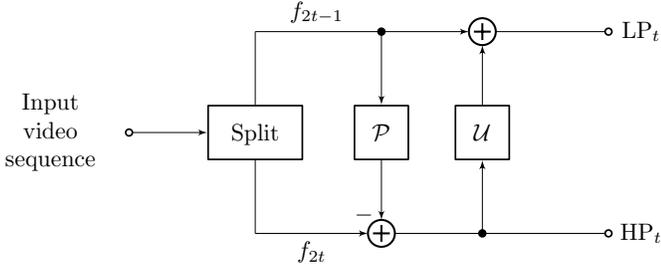}
	\caption{Block diagram of the lifting structure.}
	\label{fig:WL}
	\vspace{-2ex}
\end{figure}

The basic concept of 3D-SBC, as shown in Fig.~\ref{fig:data}, comprises a one-dimensional wavelet transform in temporal direction followed by multiple spatial wavelet transforms~\cite{4317622}. This work focuses on the temporal decomposition, while the spatial decompositions as well as the final wavelet coefficient coding are jointly implemented by the JPEG\,2000 volume coder. The temporal wavelet transform is realized by the so-called lifting structure, which is a factorized representation of the discrete wavelet transform~\cite{Sweldens1995}. As can be seen in Fig.~\ref{fig:WL}, the lifting structure comprises three steps - splitting, predicting, and updating - and results in two subbands. More precisely, the input video sequence is split into even- and odd-indexed frames $f_{2t}$ and $f_{2t-1}$. Then, the even frames are predicted from the odd frames by a prediction operator $\mathcal{P}$. The predicted frames $\mathcal{P}(f_{2t-1})$ are then subtracted from the even frames resulting in the highpass subband. Afterwards, the highpass frames are filtered by an updated operator $\mathcal{U}$ and are added to the odd frames resulting in the lowpass subband. Since the lowpass subband is very similar to the original sequence, it serves as a temporal base layer. The highpass subband contains the residual data and corresponds to the temporal enhancement layer. 
To avoid blurring and ghosting artifacts, motion compensation (MC) methods can be incorporated into the lifting structure, which is then called MCTF~\cite{334985}. Thereby, the prediction operator $\mathcal{P}$ is realized by the warping operator $\mathcal{W}_{2t-1\rightarrow2t}$. For the Haar wavelet filters, the highpass frames $\text{HP}_t$ are calculated by
\begin{equation}
\text{HP}_t = f_{2t} - \lfloor\mathcal{W}_{2t-1\rightarrow2t}(f_{2t-1})\rfloor.
\label{eq:P}
\end{equation} 
To guarantee lossless reconstruction, rounding operators are applied to achieve an integer-to-integer transform~\cite{647983}. In the update step, the motion compensation has to be inverted by switching the direction of the warping operator~\cite{1257390}. For the Haar wavelet filters the lowpass frames $\text{LP}_t$ are calculated by
\begin{equation}
\text{LP}_t = f_{2t-1} + \bigg\lfloor\frac{1}{2}\mathcal{W}_{2t\rightarrow2t-1}(\text{HP}_{t})\bigg\rfloor.
\label{eq:U}
\end{equation} 
The adaption of the transform to the signal forces a higher quality of the lowpass subband and a better energy compaction. 

\begin{figure}
	\includegraphics[width=0.98\columnwidth]{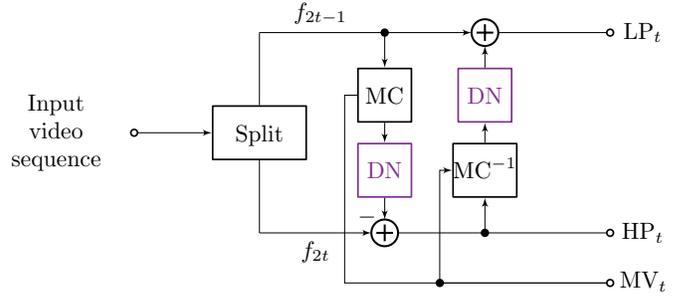}
	\caption{Block diagram of the motion compensated lifting structure containing the denoising filters (DN) of WLDPU.}
	\label{fig:WLDPU}
	\vspace{-2ex}
\end{figure}

However, for noisy CT data the coding efficiency suffers from applying MCTF. Due to the image acquisition process of CT scanners, correlated noisy structures can appear within the received image data. By warping the frames according to equations~(\ref{eq:P}) and~(\ref{eq:U}), these noisy structures get uncorrelated and the corresponding variances will add up. Consequently, the required rate for MCTF increases compared to uncompensated 3D-SBC. 
The use of spatial filtering for improved video coding has been analyzed in~\cite{1146632} and since then has been implemented successfully by several in-loop filters for hybrid coding schemes with motion compensating prediction~\cite{Norkin2012,6093807}. The application of denoising filters to motion compensated wavelet lifting has been investigated in~\cite{Lanz2018} and~\cite{lnt2019-37}.
The method, which has been developed, is called WLDPU. A block diagram of this approach is shown in Fig.~\ref{fig:WLDPU}. It can be seen that the motion compensation operations in the prediction and update steps are followed by a denoising (DN) filter, respectively. Thereby, the undesired noise increase shall be reduced leading to a significantly better compression ratio. However, the quality of the lowpass subband is also affected. The experiments in~\cite{lnt2019-37} have shown that the rate-distortion performance strongly depends on the filter strength of the specific denoising filters. So far, this value has been chosen equally for both denoising filters in advance and applied to all frames of an input video sequence.

\section{Rate-Distortion Optimization for WLDPU}
\label{sec:RDO}

As already mentioned, the performance of WLDPU depends on the filter strength, which is described by $h$. As an input parameter for both denoising filters, the filter strength is calculated by
\begin{equation}
h = \xi\cdot \sigma_n^2,
\label{eq:f}
\end{equation}
where $\sigma_n^2$ describes the noise variance of the input image and $\xi$ denotes an arbitrary noise parameter. By varying $\xi$, the strength of denoising can be controlled in order to improve the compression efficiency. This assumption is based on~\cite{lnt2011-36}, where in-loop denoising of reference frames has been proposed. $\sigma_n^2$ is calculated by a low-complexity algorithm~\cite{IMMERKAER1996300} at encoder and decoder side.

The novel idea of optimizing the selection of the filter strength is based on finding an appropriate noise parameter $\xi$ for the prediction and update step, respectively, considering a certain cost criterion. Therefore, the Lagrangian cost function for an arbitrary signal $s$ can be formulated by
\begin{equation}
C(s) = D(s) + \lambda R(s).	
\label{eq:RDO}
\end{equation}
By minimizing this cost function for every lifting step, the best matching $\xi_\text{P}$ and $\xi_\text{U}$ for a certain Lagrangian parameter $\lambda$ can be found. Thereby, the distortion $D(s)$ is calculated by the MSE of the corresponding lowpass coefficients $\text{LP}_t$ compared to the input frames according to~\cite{Lanz2018}. The rate $R(s)$ comprises the entire file size for transmitting the corresponding lowpass and highpass frames as well as the required motion vectors ($\text{MV}_t$) for lossless reconstruction at the decoder side.

The search strategy, which is applied to find the best matching $\xi_\text{P}$ and $\xi_\text{U}$ for one single lifting step, is based on a gradient descent approach~\cite{sen_kumar_2019}. Thereby, the search algorithm is initialized with $\xi_\text{P}{=}\xi_\text{U}{=}0$, which means no denoising at all. In the next iteration, $\xi_\text{P}$ and $\xi_\text{U}$ are successively incremented by $1$ and evaluated for the Lagrangian cost function. 
\begin{figure}
	\centering
	\resizebox{0.9\columnwidth}{!}{%
		\input{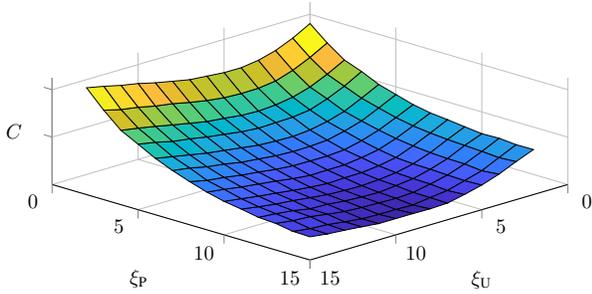}}
	\caption{Evaluation of the cost function $C(\cdot)$ for the first lifting step of the first sequence from the CT1 test data set, showing the behavior for increasing noise parameters $\xi_\text{P}$ and $\xi_\text{U}$.}
	%	\caption{Search strategy to find the best matching noise parameters $\xi_\text{P}$ and $\xi_\text{U}$.}
	\label{fig:search}
	\vspace{-2ex}
\end{figure}
%Since the simulations in~\cite{lnt2019-37} have shown that with increasing filter strengths the rate as well as the quality are monotonically decreased, a top-to-bottom search strategy can be used to minimize equation~(\ref{eq:RDO}). 
%The Lagrangian cost function is evaluated for increasing values of $\xi$ 
This procedure is continued as long as the minimum costs of the next iteration $i{+}1$ are lower or equal than the minimum costs of the current iteration $i$
\begin{align*}
D(\text{LP}_{t,i}) + \lambda R(\text{LP}_{t,i}+\text{HP}_{t,i}+&\text{MV}_{2t\rightarrow 2t-1,i}) \leq \numberthis \\
D(\text{LP}_{t,i+1}) + \lambda R(\text{LP}_{t,i+1}+&\text{HP}_{t,i+1}+\text{MV}_{2t\rightarrow 2t-1,i+1}).
\end{align*}
Obviously, gradient descent does not guarantee to find the global minimum in general. However, the simulations in~\cite{lnt2019-37} have shown that with increasing values of $\xi$ the rate as well as the quality are monotonically decreasing. This is also illustrated by Fig.~\ref{fig:search}, showing the evaluation of the cost function for one lifting step of the first sequence from the CT test data set. Therefore, a gradient descent approach is assumed to be sufficient to find the global minimum.

The Lagrange multiplier $\lambda$ can be any positive value and controls the balance of rate and distortion. Using a fixed value of $\lambda$ as input parameter for the encoder, the desired distortion of the lowpass subband shall be adjusted by self-controlled selection of optimum values for $\xi_\text{P}$ and $\xi_\text{U}$. For large values of $\lambda$, the rate is decreased and for small values, the distortion is decreased.
For simulation, eight different values of $\lambda$ will be tested, following the geometric series
\begin{equation}
\lambda_{n+1} = \lambda_n \cdot 3,
\end{equation}
starting with $\lambda_0=0.05$. The reason for this series will be given in Section~\ref{sec:results}. Experimental results have shown that the interval $[\lambda_0,\lambda_4]$ can be used for achieving high-quality lowpass subbands, while the interval $[\lambda_4,\lambda_7]$ results in low to medium quality but in a higher overall compression. Due to the gradient descent approach for searching the optimum values for $\xi_\text{P}$ and $\xi_\text{U}$, the computational complexity increases with higher values of $\lambda$.
%To make the choice of $\lambda$ more intuitive, we use the \mbox{$\lambda$-QP relation}, which is well-known from state-of-the-art coders like H.264/AVC~\cite{ITU-T-H-264} and HEVC
%\begin{equation}
%	\lambda = \frac{2}{3}\cdot 2^{\frac{\text{QP}-12}{3}}.
%\end{equation}
%This allows for facilitating the selection of $\lambda$ with respect to a global rate-distortion ratio specified by QP. While the values $\text{QP}{=}2,7,12,17,22$ can be used for achieving high-quality lowpass subbands, the values $\text{QP}{=}22,27,32,37$ result in low to medium quality but in a higher overall compression.

Obviously, the proposed method, which will be abbreviated by $\text{WLDPU}_\text{RDO}$, is signal-dependent. In every lifting step a specific filter strength for the prediction and update denoising filters, respectively, is calculated. Therefore, $\xi_\text{P}$ and $\xi_\text{U}$ have to be transmitted as side information to guarantee lossless reconstruction at decoder side. 
%Instead of transmitting the exact values of the filter strengths, it is enough to transmit just the noise parameters $\xi_\text{pred}$ and $\xi_\text{upd}$. 
%According to equation (\ref{eq:f}), $h$ can be calculated from $\xi$ and the corresponding noise variance, which is still the same than before. 
%Since only integer values are used for varying $\xi$, entropy coding is very cheap. 
In this work, this is done by multiple-context adaptive arithmetic coding~\cite{Witten:1987}. 

\section{Simulation Results}
\label{sec:results}
\begin{table}
	\centering	
	\caption{For simulation, two \mbox{3D+t} CT data sets, comprising 127 and 200 temporal sequences, as well as 14 independent temporal sequences from a MRI device are used.}
	\resizebox{0.98\columnwidth}{!}{%
	\label{tab:data}
	\begin{tabular}{l|rrrr|c}
		\toprule
		& \multicolumn{1}{c}{x} & \multicolumn{1}{c}{y} & \multicolumn{1}{c}{z} & \multicolumn{1}{c|}{t} & \multicolumn{1}{l}{$\#$ Temporal sequences} \\ \midrule
		CT1  & 512                   & 512                   & 127                   & 10                     & 127                                             \\
		CT2  & 512                   & 512                   & 200                   & 10                     & 200                                            \\
		MRI & 144                   & 192                   & 1                     & 25                     & 14                                             \\ \bottomrule
	\end{tabular}}
	\vspace{-2ex}
\end{table}

\begin{figure*}
	\centering
	\resizebox{0.98\textwidth}{!}{%
		\input{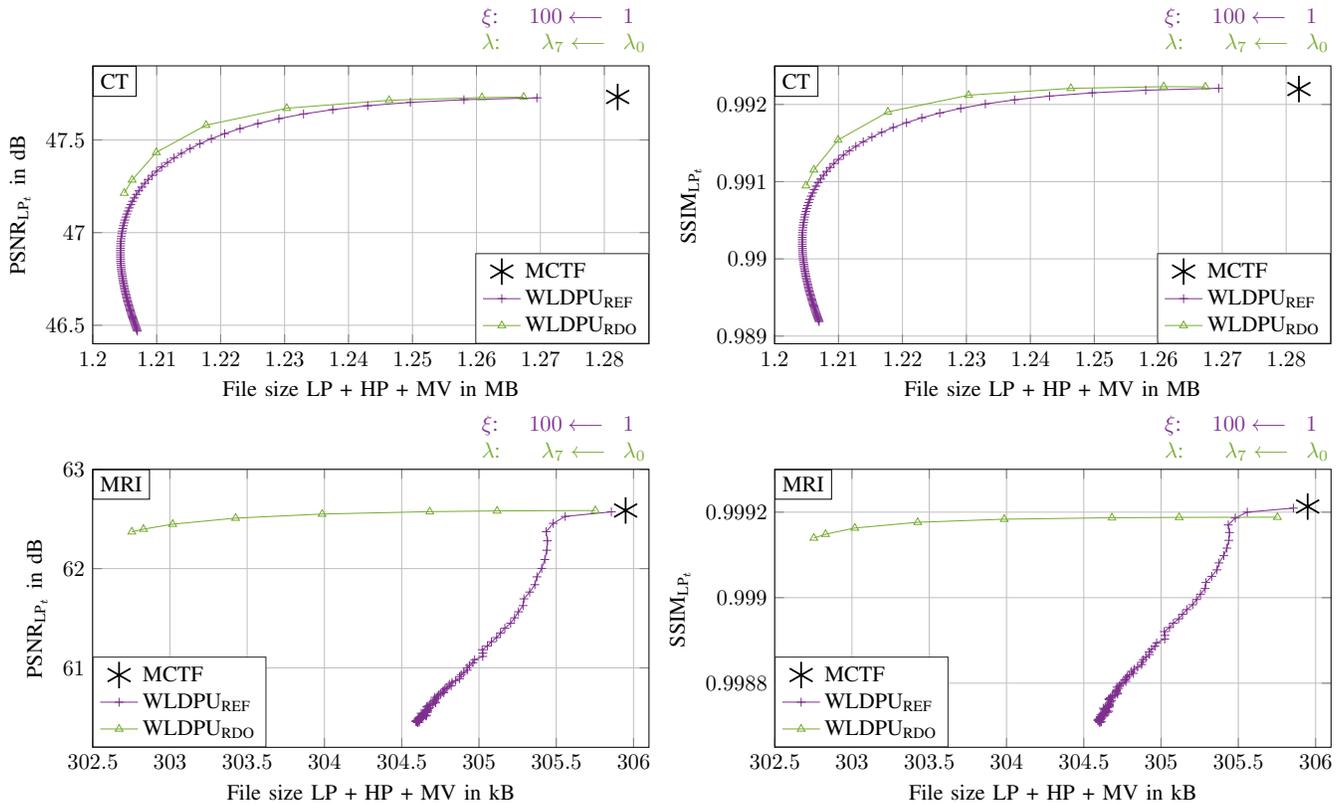}}
	\caption{$\text{PSNR}_{\text{LP}_t}$ (left) and $\text{SSIM}_{\text{LP}_t}$ (right) results, respectively, plotted against the overall file size comparing $\text{WLDPU}_\text{REF}$ to $\text{WLDPU}_\text{RDO}$ for the CT and MRI data. The arrows above the diagrams indicate the direction for increasing values of $\xi$ and $\lambda$, respectively.
		%	For better presentation, only every eighth value for $\xi$ is plotted. 
		Results are averaged over all sequences of the CT and MRI data sets.}
	\label{fig:RDO}
	\vspace{-2ex}
\end{figure*}
All simulations are evaluated on two 3D+$t$ CT data sets\footnote{The CT volume data sets were kindly provided by Siemens Healthineers.} as well as on a 2D+$t$ MRI data set. As summarized in Table~\ref{tab:data}, both CT data sets comprise 10 time steps and have a spatial resolution of $512{\times}512$, which is typical for CT data, while the MRI sequences offer a spatial resolution of $192\times144$. While the first CT data set consists of $127$ slices, the second CT data set comprises $200$ slices, which results in $327$ single temporal CT sequences. The MRI data set comprises $14$ independent temporal sequences each with $25$ time steps. All data sets have a bit depth of $12$ bits per sample and show a beating heart over time. 

The temporal WT is implemented according to Section~\ref{sec:WL} with a mesh-based MC~\cite{8066388}, where a grid size of $8{\times}8$ pixels is used. For the spatial WT performed by the JPEG\,2000 coder, four decomposition levels inside the OpenJPEG~\cite{openjpeg} implementation are used. The MVs are encoded using the QccPack library~\cite{fowler2000qccpack}. 
In accordance to the cost criterion introduced in Section~\ref{sec:RDO}, the quality of the lowpass subband is evaluated by the $\text{PSNR}_{\text{LP}_t}$ metric from~\cite{Lanz2018}. To validate the results, the modified structural similarity index $\text{SSIM}_{\text{LP}_t}$, as also proposed in~\cite{Lanz2018}, is additionally evaluated. These metrics both consider not only the similarity of the lowpass subband to the odd-indexed frames but also to the even-indexed frames. 
%For denoising, Block Matching and 3-D Filtering~(BM3D)~\cite{Dabov2007} is applied in this work. 
Besides simple Gaussian filtering, also more complex filters have been investigated in~\cite{Lanz2018} for compensated wavelet lifting with denoising filters. 
%In the framework of WLDPU, BM3D filtering performs best, as has been shown in~\cite{Lanz2016}.
Since Block Matching and 3-D Filtering~(BM3D)~\cite{4271520} turned out to perform best in the framework of WLDPU, it is also applied in this work to perform the denoising operations in the prediction and update steps, respectively. 
%The reference implementation will further be abbreviated by $\text{WLDPU}_\text{REF}$.
As already mentioned above, for the proposed $\text{WLDPU}_\text{RDO}$, eight different values of $\lambda$ will be used for evaluation. The results will be compared to the reference implementation of WLDPU from~\cite{lnt2019-37}, called $\text{WLDPU}_\text{REF}$, where the noise parameters are chosen equally for all denoising operations performed for each sequence. Thereby, $\xi=\xi_\text{P}=\xi_\text{U}$ will be varied in a range of integer values from $1$ to $100$. The results are averaged over all $327$ CT sequences and all $14$ MRI sequences, respectively.

\begin{table}
	\centering	
	\caption{Average bit rate savings for $\text{WLDPU}_\text{RDO}$ compared to $\text{WLDPU}_\text{REF}$. The ranges for $\lambda_n$ with HQ and LQ are considered separately. Results are averaged over all sequences of the CT and MRI data sets.}
	\resizebox{0.8\columnwidth}{!}{%
		\begin{tabular}{l|cc}\toprule
			BD-rate &  HQ  &  LQ \\ 
			%		         & QP=2,7,12,17,22 & QP = 22,27,32,37\\ \midrule
			& $\lambda_0,\lambda_1,\lambda_2,\lambda_3,\lambda_4$ & $\lambda_4,\lambda_5,\lambda_6,\lambda_7$\\ \midrule
			CT   & -0.73\,\% & -0.32\,\%\\  
			MRI  & -0.58\,\% & -0.80\,\%\\  \bottomrule
		\end{tabular}%
	}	
	\label{tab:BD}
	\vspace{-2ex}
\end{table}
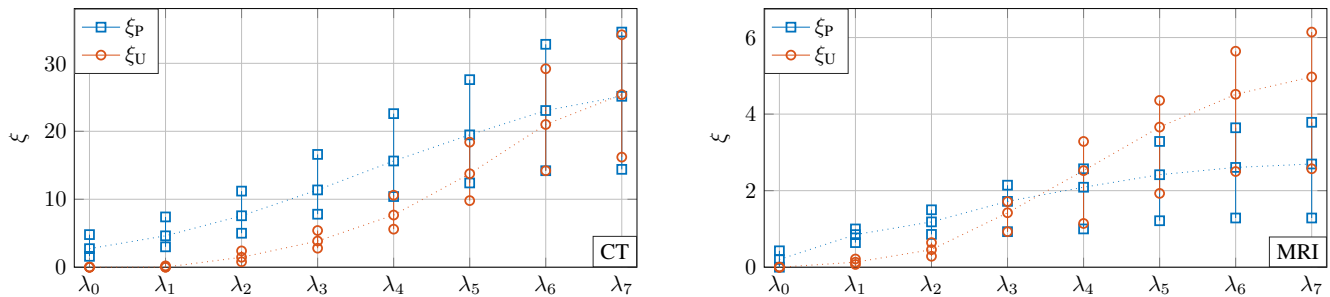
\begin{figure*}
	\centering
	\resizebox{0.98\textwidth}{!}{%
		\definecolor{mycolorred}{rgb}{0.85000,0.32500,0.09800}%
\definecolor{mycolorblue}{rgb}{0.00000,0.44700,0.74100}%
\definecolor{mycolorpurple}{rgb}{0.49400,0.18400,0.55600}%
\definecolor{mycolorgreen}{rgb}{0.46600,0.67400,0.18800}%
\definecolor{mycolorblack}{rgb}{0.00,0.00,0.00}%
\definecolor{mycoloryellow}{rgb}{1,0.65,0}%
\begin{tikzpicture}

\begin{groupplot}[group style={group name = group,group size=2 by 1, horizontal sep = 2cm, vertical sep=1.1cm}]

\nextgroupplot[%
width=\columnwidth,
height=0.46\columnwidth,
scale only axis,
xtick={2,7,12,17,22,27,32,37},
xticklabels={$\lambda_0$,$\lambda_1$,$\lambda_2$,$\lambda_3$,$\lambda_4$,$\lambda_5$,$\lambda_6$,$\lambda_7$},
%x tick label style={rotate=45,anchor=east},
xmajorgrids,
xmin=1,
xmax=38,
ymin=0,
ylabel={$\xi$},
ymajorgrids,
axis background/.style={fill=white},
title={CT},
every axis title/.style={draw,fill=white,at={(1,0)},anchor=south east,draw=white!15!black},
legend style={legend cell align=left,at={(0,1)},anchor=north west,draw=white!15!black}
]

\addplot [color=mycolorblue,solid,mark=square,mark options={thick,solid}]
table[row sep=crcr]{%
	2	1.6\\
	2	2.7333\\
	2	4.8\\
};\addlegendentry{$\xi_\text{P}$}

\addplot [color=mycolorred,solid,mark=o,mark options={thick,solid}]
table[row sep=crcr]{%
	2	0\\
	2	0\\
	2	0\\
};
\addlegendentry{$\xi_\text{U}$}

\addplot [color=mycolorblue,solid,mark=square,mark options={thick,solid}]
table[row sep=crcr]{%
	7	3\\
	7	4.6210\\
	7	7.4\\
};

\addplot [color=mycolorblue,solid,mark=square,mark options={thick,solid}]
table[row sep=crcr]{%
	12	5\\
	12	7.5710\\
	12	11.2\\
};

\addplot [color=mycolorblue,solid,mark=square,mark options={thick,solid}]
table[row sep=crcr]{%
	17	7.8\\
	17	11.3728\\
	17	16.6\\
};

\addplot [color=mycolorblue,solid,mark=square,mark options={thick,solid}]
table[row sep=crcr]{%
	22	10.4\\
	22	15.6321\\
	22	22.6\\
};

\addplot [color=mycolorblue,solid,mark=square,mark options={thick,solid}]
table[row sep=crcr]{%
	27	12.4\\
	27	19.4778\\
	27	27.6\\
};

\addplot [color=mycolorblue,solid,mark=square,mark options={thick,solid}]
table[row sep=crcr]{%
	32	14.2\\
	32	23.0574\\
	32	32.8\\
};

\addplot [color=mycolorblue,solid,mark=square,mark options={thick,solid}]
table[row sep=crcr]{%
	37	14.4\\
	37	25.1500\\
	37	34.6\\
};

\addplot [color=mycolorred,solid,mark=o,mark options={thick,solid}]
table[row sep=crcr]{%
	7	0\\
	7	0.0154\\
	7	0.2\\
};

\addplot [color=mycolorred,solid,mark=o,mark options={thick,solid}]
table[row sep=crcr]{%
	12	0.8\\
	12	1.4617\\
	12	2.4\\
};

\addplot [color=mycolorred,solid,mark=o,mark options={thick,solid}]
table[row sep=crcr]{%
	17	2.8\\
	17	3.8512\\
	17	5.4000\\
};

\addplot [color=mycolorred,solid,mark=o,mark options={thick,solid}]
table[row sep=crcr]{%
	22	5.6\\
	22	7.6747\\
	22	10.6000\\
};

\addplot [color=mycolorred,solid,mark=o,mark options={thick,solid}]
table[row sep=crcr]{%
	27	9.8\\
	27	13.7562\\
	27	18.4000\\
};

\addplot [color=mycolorred,solid,mark=o,mark options={thick,solid}]
table[row sep=crcr]{%
	32	14.2\\
	32	21.0037\\
	32	29.2000\\
};

\addplot [color=mycolorred,solid,mark=o,mark options={thick,solid}]
table[row sep=crcr]{%
	37	16.2\\
	37	25.4000\\
	37	34.2\\
};

\addplot [color=mycolorblue,dotted]
table[row sep=crcr]{%
	2	2.7333\\
	7	4.6210\\
	12	7.5710\\
	17	11.3728\\
	22	15.6321\\
	27	19.4778\\
	32	23.0574\\
	37	25.1500\\
};

\addplot [color=mycolorred,dotted]
table[row sep=crcr]{%
	2	0\\
	7	0.0154\\
	12	1.4617\\
	17	3.8512\\
	22	7.6747\\
	27	13.7562\\
	32	21.0037\\
	37	25.4000\\
};

\nextgroupplot[%
width=\columnwidth,
height=0.46\columnwidth,
scale only axis,
xtick={2,7,12,17,22,27,32,37},
xticklabels={$\lambda_0$,$\lambda_1$,$\lambda_2$,$\lambda_3$,$\lambda_4$,$\lambda_5$,$\lambda_6$,$\lambda_7$},
%x tick label style={rotate=45,anchor=east},
xmajorgrids,
xmin=1,
xmax=38,
ymin=0,
ylabel={$\xi$},
ymajorgrids,
axis background/.style={fill=white},
title={MRI},
every axis title/.style={draw,fill=white,at={(1,0)},anchor=south east,draw=white!15!black},
legend style={legend cell align=left,at={(0,1)},anchor=north west,draw=white!15!black}
]

\addplot [color=mycolorblue,solid,mark=square,mark options={thick,solid}]
table[row sep=crcr]{%
	2	0.428571428571429\\
	2	0.202380952380952\\
	2	0\\
};\addlegendentry{$\xi_\text{P}$}

\addplot [color=mycolorred,solid,mark=o,mark options={thick,solid}]
table[row sep=crcr]{%
	2	0\\
	2	0\\
	2	0\\
};
\addlegendentry{$\xi_\text{U}$}

\addplot [color=mycolorblue,solid,mark=square,mark options={thick,solid}]
table[row sep=crcr]{%
	7	1\\
	7	0.851190476190476\\
	7	0.642857142857143\\
};

\addplot [color=mycolorblue,solid,mark=square,mark options={thick,solid}]
table[row sep=crcr]{%
	12	1.50000000000000\\
	12	1.18452380952381\\
	12	0.857142857142857\\
};

\addplot [color=mycolorblue,solid,mark=square,mark options={thick,solid}]
table[row sep=crcr]{%
	17	2.14285714285714\\
	17	1.72023809523810\\
	17	0.928571428571429\\
};

\addplot [color=mycolorblue,solid,mark=square,mark options={thick,solid}]
table[row sep=crcr]{%
	22	2.57142857142857\\
	22	2.08928571428571\\
	22	1\\
};

\addplot [color=mycolorblue,solid,mark=square,mark options={thick,solid}]
table[row sep=crcr]{%
	27	3.28571428571429\\
	27	2.41666666666667\\
	27	1.21428571428571\\
};

\addplot [color=mycolorblue,solid,mark=square,mark options={thick,solid}]
table[row sep=crcr]{%
	32	3.64285714285714\\
	32	2.60714285714286\\
	32	1.28571428571429\\
};

\addplot [color=mycolorblue,solid,mark=square,mark options={thick,solid}]
table[row sep=crcr]{%
	37	3.78571428571429\\
	37	2.69642857142857\\
	37	1.28571428571429\\
};

\addplot [color=mycolorred,solid,mark=o,mark options={thick,solid}]
table[row sep=crcr]{%
	7	0.214285714285714\\
	7	0.130952380952381\\
	7	0.0714285714285714\\
};

\addplot [color=mycolorred,solid,mark=o,mark options={thick,solid}]
table[row sep=crcr]{%
	12	0.642857142857143\\
	12	0.458333333333333\\
	12	0.285714285714286\\
};

\addplot [color=mycolorred,solid,mark=o,mark options={thick,solid}]
table[row sep=crcr]{%
	17	1.71428571428571\\
	17	1.42261904761905\\
	17	0.928571428571429\\
};

\addplot [color=mycolorred,solid,mark=o,mark options={thick,solid}]
table[row sep=crcr]{%
	22	3.28571428571429\\
	22	2.52380952380952\\
	22	1.14285714285714\\
};

\addplot [color=mycolorred,solid,mark=o,mark options={thick,solid}]
table[row sep=crcr]{%
	27	4.35714285714286\\
	27	3.66071428571429\\
	27	1.92857142857143\\
};

\addplot [color=mycolorred,solid,mark=o,mark options={thick,solid}]
table[row sep=crcr]{%
	32	5.64285714285714\\
	32	4.51785714285714\\
	32	2.50000000000000\\
};

\addplot [color=mycolorred,solid,mark=o,mark options={thick,solid}]
table[row sep=crcr]{%
	37	6.14285714285714\\
	37	4.97023809523810\\
	37	2.57142857142857\\
};

\addplot [color=mycolorblue,dotted]
table[row sep=crcr]{%
	2	0.202380952380952\\
	7	0.851190476190476\\
	12	1.18452380952381\\
	17	1.72023809523810\\
	22	2.08928571428571\\
	27	2.41666666666667\\
	32	2.60714285714286\\
	37	2.69642857142857\\
};

\addplot [color=mycolorred,dotted]
table[row sep=crcr]{%
	2	0\\
	7	0.130952380952381\\
	12	0.458333333333333\\
	17	1.42261904761905\\
	22	2.52380952380952\\
	27	3.66071428571429\\
	32	4.51785714285714\\
	37	4.97023809523810\\
};

\end{groupplot}
\end{tikzpicture}}
	\caption{Variance of selected values for $\xi_\text{P}$ and $\xi_\text{U}$ depending on the value of $\lambda$. Results are averaged over all sequences of the CT and MRI data sets.}
	\label{fig:variances}
	\vspace{-2ex}
\end{figure*}

Fig.~\ref{fig:RDO} shows the performance of the two different approaches. On the left side, the \mbox{$\text{PSNR}_{\text{LP}_t}$-rate} results for the CT and MRI data, respectively, are provided, while the \mbox{$\text{SSIM}_{\text{LP}_t}$-rate} results are given at the right side of Fig.~\ref{fig:RDO}. Apart from a small offset regarding the $\text{SSIM}_{\text{LP}_t}$ results of the MRI data, which can be explained by using another metric for evaluation than for optimization, both metrics are showing the same behavior, which is why all further analyses are exclusively done for $\text{PSNR}_{\text{LP}_t}$. 
%The offset can be explained by the calculation of the distortion used within the evaluation of the cost function. Since the $\text{SSIM}_{\text{LP}_t}$ metric is not based on the MSE value used within $\text{PSNR}_{\text{LP}_t}$ and the RDO, the evaluation does not inherently has to give optimum performance results.
The performance of MCTF without any filtering steps is shown in black, while the purple curves provide the results for $\text{WLDPU}_\text{REF}$. The purple arrows above the diagrams indicate the direction for increasing values of $\xi$. For the CT data, it is approved that for higher filter strengths the compression ratio is enhanced, but the quality in terms of $\text{PSNR}_{\text{LP}_t}$ suffers significantly for noise parameters \mbox{$\xi\geq25$}. However, up to this point the performance of $\text{WLDPU}_\text{REF}$ is already promising. In contrast, for the MRI sequences, the quality drops almost linearly with the rate, which is disadvantageous for telemedicine applications, where a high data fidelity is indispensable. The achieved bit rate savings are also less convincing.
%Markers are plotted only for $\xi=[1,2,3,7,11,17,23,25]$.
The green curves show the performance of the proposed $\text{WLDPU}_\text{RDO}$. The corresponding arrows above the diagrams show the direction for increasing input parameters $\lambda$.
For the CT data, it can be observed that the abrupt loss regarding $\text{PSNR}_{\text{LP}_t}$ can be avoided by the novel approach. This is caused by forcing the proposed algorithm to stop if the cost criterion is not fulfilled anymore.
%inverse behavior of the reference implementation for filter strengths greater than $25$ can be avoided by the novel approach.  
%Due to too strong filtering in the prediction step of $\text{WLDPU}_\text{REF}$, the inverse MC in the update step does not give reasonable results. So far, this resulted in a fast drop of the quality and an increase of the overall rate. This effect is now prevented for $\text{WLDPU}_\text{RDO}$ by forcing the proposed algorithm to stop, if the cost criterion is not fulfilled anymore. 
For the MRI data, the advantage of the proposed $\text{WLDPU}_\text{RDO}$ is even more dominant. The linear decrease of the quality can be prevented and additional bit rate savings can be achieved compared to $\text{WLDPU}_\text{REF}$.

Apart from that, it can be observed why the geometric distribution regarding the different values of $\lambda$ is suggested. From $\lambda_0$ to $\lambda_6$, almost equidistant rate differences can be achieved. For $\lambda_7$, no significant improvement is reached anymore, indicating a saturation regarding the maximum possible bit rate savings resulting from $\text{WLDPU}_\text{RDO}$. 
%Of course, every other positive value of $\lambda$ would be possible to achieve a rate-distortion optimization within the framework of $\text{WLDPU}_\text{RDO}$.

For quantitative evaluation, the average bit rate savings of the proposed method compared to $\text{WLDPU}_\text{REF}$ in terms of Bj\o{}ntegaard-Delta rate\,(BD-rate)~\cite{10016799583} are calculated. The results are provided in Table~\ref{tab:BD}. Thereby, the considered $\lambda$ ranges are separated into high quality\,(HQ) of the lowpass subband and low to medium quality\,(LQ). As can be seen from Tab.~\ref{tab:BD}, for the CT data, the RDO in the context of WLDPU achieves average bit rate savings of $0.73$\,\% for the HQ range and $0.32$\,\% for the LQ range. For the MRI data, average bit rate savings of $0.58$\,\% for the HQ range and $0.80$\,\% for the LQ range can be achieved.

To understand where these bit rate savings come from, Fig.~\ref{fig:variances} shows the variances of self-selected noise parameters $\xi_\text{P}$ and $\xi_\text{U}$ regarding each tested input parameter $\lambda$. Thereby, the minimum and maximum values after averaging over the CT and MRI data sets, respectively, are presented for every $\lambda$. The associated mean values are connected by dotted lines.
As expected, the absolute values of selected noise parameters are increasing with rising values of $\lambda$, allowing for a stronger denoising and therefore, to higher bit rate savings. Furthermore it can be observed that with higher $\lambda$ values the variance of selected noise parameters becomes larger. This means that for achieving higher bit rate savings and preserving the data fidelity, a smart choice of $\xi_\text{P}$ and $\xi_\text{U}$ is of huge importance. This is mainly caused by the content dependency of the underlying sequences. Especially for the MRI data set the related gains are significant, as the rate-distortion plots in Fig.~\ref{fig:RDO} have already shown.
%This is the reason, why the BD-PSNR for HQ is very low. Since the range of selected values is very small, the advantage of individually selecting noise parameters is less prominent. However, with higher $\lambda$ values the variance of selected noise parameters becomes larger indicating that the choice of $\xi_\text{P}$ and $\xi_\text{U}$ is highly dependent on the content of the underlying sequence. 
%This enables the RDO to outperform the previous setup. 
Additionally, the diagrams show that for very low values of $\lambda$, the denoising filter in the update step is nearly turned off. This allows for saving computational complexity by omitting the denoising filter in the update step, when a HQ lowpass subband is desired. For the CT data sets, the update filter was omitted in approximately $25\%$ of all denoising processes across all values of $\lambda$, while for the MRI sequences, it was omitted in almost $60\%$.

To classify the performance of the proposed $\text{WLDPU}_\text{RDO}$ and show the relevance of compensated wavelet lifting for medical data, in Table.~\ref{tab:comp} a comparison between the compression efficiency of HEVC, \mbox{3D-SBC}, MCTF, $\text{WLDPU}_\text{REF}$, and $\text{WLDPU}_\text{RDO}$ is provided. Additionally, single image coding by JPEG-LS, still representing the most widespread compression scheme for medical content, is taken into account. For HEVC, the \textit{Lowdelay Main RExt} configuration and lossless mode within the test model HM-16.16 are employed. The wavelet-based compression schemes are evaluated using the coder specifications introduced above. For $\text{WLDPU}_\text{RDO}$, $\lambda_7$ is applied, while $\text{WLDPU}_\text{REF}$ is evaluated for $\xi=25$ and $\xi=4$ for the CT and MRI data sets, respectively, to achieve the same quality of the lowpass subband in terms of $\text{PSNR}_{\text{LP}_t}$ as for $\text{WLDPU}_\text{RDO}$. Consequently, several observations can be made. First of all, JPEG-LS is outperformed by all other coding schemes. This shows the large potential regarding inter-prediction for efficient lossless coding of medical content. Secondly, HEVC performs less efficient than all wavelet-based coding schemes for CT data. Considering MRI data, 3D-SBC also outperforms HEVC. This may be caused by the fact that HEVC has mainly been developed for lossy compression of video content from consumer applications. As a side benefit, the wavelet-based coding schemes inherently provide a scalable bit stream. Obviously, the visual quality of the temporal lowpass subband in terms of $\text{PSNR}_{\text{LP}_t}$ resulting from uncompensated 3D-SBC is significantly lower than for MCTF, which however, leads to an increasing file size. To compensate this, $\text{WLDPU}_\text{REF}$ can be applied, reducing the required rate as well as preserving the data fidelity at approximately the same level as MCTF. 
The small losses of $0.52$\,dB and $0.22$\,dB regarding the CT and MRI data sets, respectively, are barely visible due to the overall high level of $\text{PSNR}_{\text{LP}_t}$, while the achieved bit rate savings are proportionally high compared to MCTF.
These bit rate savings can be increased further by employing the novel approach of $\text{WLDPU}_\text{RDO}$.
\begin{table}
	\centering
	\caption{Comparison regarding the compression efficiency of several coding schemes for CT and MRI data sets.}
	\label{tab:comp}
	\resizebox{0.98\columnwidth}{!}{%
	\begin{tabular}{l|rr||rr}\toprule
		        & \multicolumn{2}{c||}{CT} & \multicolumn{2}{c}{MRI}               \\ \hline
		        & \multicolumn{1}{c|}{File size}   & \multicolumn{1}{c||}{$\text{PSNR}_{\text{LP}_t}$ } & \multicolumn{1}{c|}{File size}  & \multicolumn{1}{c}{$\text{PSNR}_{\text{LP}_t}$ }\\
		        & \multicolumn{1}{c|}{in kB}   & \multicolumn{1}{c||}{in dB} & \multicolumn{1}{c|}{in kB}  & \multicolumn{1}{c}{in dB}\\\midrule
		HEVC-LD & \multicolumn{1}{r|}{1462.19}        &           & \multicolumn{1}{r|}{297.01}  &       \\
		3D-SBC  & \multicolumn{1}{r|}{1201.39}        & 42.85     & \multicolumn{1}{r|}{295.47}  & 59.57 \\
		MCTF    & \multicolumn{1}{r|}{1312.85}        & 47.73     & \multicolumn{1}{r|}{305.95}  & 62.59 \\
		$\text{WLDPU}_\text{REF}$   & \multicolumn{1}{r|}{1235.79}        & 47.21     & \multicolumn{1}{r|}{305.43}  & 62.37 \\
		$\text{WLDPU}_\text{RDO}$   & \multicolumn{1}{r|}{1233.84}        & 47.21     & \multicolumn{1}{r|}{302.75}  & 62.37 \\		
		JPEG-LS & \multicolumn{1}{r|}{3631.39}        &           & \multicolumn{1}{r|}{1023.34} &      \\\bottomrule
	\end{tabular}}
	\vspace{-2ex}
\end{table}

\section{Conclusion}
\label{sec:conclusion}
Scalable lossless coding is a very important aspect for telemedicine applications. MCTF describes a fully scalable coding framework providing a high quality lowpass subband, which can be used as a temporal representative of the original sequence. However, the compression efficiency suffers from noisy CT and MRI data. Integrating denoising filters into the lifting steps of MCTF turned out to offer significant bit rate savings but strongly depends on the used filter parameters. In this work, a rate-distortion optimization is integrated into the existing framework of WLDPU. Thereby, the optimal filter strength for each denoising process can be selected individually depending on a single input parameter for the encoder to control the rate-distortion performance. This provides additional bit rate savings in terms of BD-rate and stabilizes the system performance considerably. Further research aims at closing the remaining gap between $\text{WLDPU}_\text{RDO}$ and \mbox{3D-SBC} to achieve an efficient scalable compression scheme, supporting a scalable representation required for telemedicine applications and preserving DICOM compatibility.

\bibliographystyle{IEEEtran}
\bibliography{Literatur.bib}

\end{document}